\documentclass[aps,prl,twocolumn,showpacs]{revtex4}
\usepackage{bbm,amsmath,amssymb,amsbsy,graphicx,times,subfigure}

\usepackage[latin1]{inputenc}
\newcommand{\R}{\mathbbm{R}}

\newcommand{\id}{\mathbbm{1}}

\renewcommand{\det}{{\rm Det}\,}
\newcommand{\gr}[1]{\boldsymbol{#1}}
\newcommand{\be}{\begin{equation}}
\newcommand{\ee}{\end{equation}}
\newcommand{\bea}{\begin{eqnarray}}
\newcommand{\eea}{\end{eqnarray}}

\newcommand{\sig}{\gr{\sigma}}
\newcommand{\eps}{\gr{\varepsilon}}

\newcommand{\eq}[1]{Eq.~(\ref{#1})}

\begin{document}
\title{Generic Entanglement and Standard Form for $\gr N$-Mode Pure Gaussian States}
\author{Gerardo Adesso}
\affiliation{Dipartimento di Fisica ``E. R. Caianiello'',
Universit\`a degli Studi di Salerno, INFN Sezione di Napoli-Gruppo
Collegato di Salerno, Via S. Allende, 84081 Baronissi (SA), Italy
\\ Centre for Quantum Computation, DAMTP, Centre for Mathematical
Sciences, University of Cambridge, Wilberforce Road, Cambridge CB3
0WA, United Kingdom}

\date{September 25, 2006}

\pacs{03.67.Mn, 03.65.Ud, 42.50.Dv}

\begin{abstract}We investigate the correlation structure of pure $N$-mode
Gaussian resources which can be experimentally generated by means of
squeezers and beam splitters, whose entanglement properties are {\em
generic}. We show that those states are specified (up to local
unitaries) by $N(N-1)/2$ parameters, corresponding to the two-point
correlations between any pair of modes.  Our construction yields a
practical scheme to engineer such generic-entangled $N$-mode pure
Gaussian states by linear optics. We discuss our findings in the
framework of Gaussian matrix product states of harmonic lattices,
raising connections with entanglement frustration and the entropic
area law.
\end{abstract}
\maketitle

\paragraph{Introduction.\,{\em ---}}

Multipartite entanglement in pure  states of many systems is a
founding property and a crucial resource for quantum information
science, yet its complete theoretical understanding is still
lacking. A basic property of entanglement is its invariance under
unitary operations performed locally on the subsystems. To describe
entanglement efficiently, is thus natural to lighten quantum systems
of the unnecessary degrees of freedom adjustable by local unitaries
(LUs), and to classify states according to {\em standard forms}
representative of LU equivalence classes \cite{linden}.
Alongside the traditional
qubit-based approach, quantum information with continuous
variables (CV) is a burgeoning field mainly spinning around the
theory and applications of entanglement in Gaussian states
\cite{review}.

In this Letter we address the question of how many physical
resources are really needed to engineer and characterize
entanglement in pure Gaussian states of an arbitrary number of
modes, up to LU operations. For states of $N \le 3$ modes, it has
been shown that such a number of minimal degrees of freedom scales
as $N(N-1)/2$ \cite{2msform,3modi}. For a higher number of modes,
however, a richer structure is achievable by pure Gaussian states,
as from the normal form of Ref. \cite{GEOF} a minimal number of
parameters given by $N(N-2)$ can be inferred \cite{private}. A
random state of $N \ge 4$ modes, selected according to the uniform
distribution over pure Gaussian states, will be thus reducible to
a form characterized by such a number of independent quantities.
However, in practical realizations of CV quantum information one
is interested in states which, once prepared with efficient
resources, still achieve an almost complete structural variety in
their multipartite entanglement properties. Such states will be
said to possess {\em generic entanglement} \cite{generic}, where
generic means practically equivalent to that of random states, but
engineered (and described) with a considerably smaller number of
degrees of freedom.


Precisely, we define as ``generic-entangled'' those Gaussian
states whose local entropies of entanglement in any single mode
are independent, and bipartite entanglements between any pair of
modes are unconstrained. Having a standard form for such $N$-mode
Gaussian states, may be in fact extremely helpful in understanding
and quantifying multipartite CV entanglement, in particular from
the theoretical point of view of entanglement sharing and monogamy
constraints \cite{contangle,3modi}, and from a more pragmatical
approach centered on using entanglement as a resource. We show
that, to achieve generic entanglement, for the global pure
$N$-mode Gaussian state it is enough to be described by a minimal
number of
 parameters (corresponding to the LU invariant
degrees of freedom) equal to $N(N-1)/2$ for any $N$, and thus much
smaller than the $2N(2N+1)/2$ of a completely general covariance
matrix. Therefore, generic entanglement appears in states which are
highly {\em not} `generic' in the sense usually attributed to the
term, {\em i.e.}~randomly picked.
Crucially, we demonstrate that generic-entangled Gaussian states
coincide with the resources typically employed in experimental
realizations of CV quantum information \cite{review}, and we provide
an optimal scheme for their state engineering.


\paragraph{Preliminaries.\,{\em ---}}

We consider a CV system consisting of $N$ canonical bosonic modes,
and described by the vector $\hat{X} = \{\hat q_1, \hat p_1, \hat
q_2, \hat p_2, \ldots, \hat q_N, \hat p_N\}$ of the field quadrature
operators, which satisfy the commutation relations $[\hat X_{i},\hat
X_j]=2i\Omega_{ij}$, with the symplectic form
$\gr\Omega=\gr\omega^{\oplus N}$ and $\gr\omega={{0\ \ 1}\choose
{-1\ 0}}$. Gaussian states (such as vacua, coherent, and squeezed
states) are defined by having a Gaussian characteristic function in
phase space \cite{review}. They are fully characterized by the first
statistical moments (arbitrarily adjustable by LUs: we will set them
to zero) and by the $2N \times 2N$ covariance matrix (CM) $\sig$ of
the second moments
 $\sigma_{ij}=\langle\{\hat{X}_i,\hat{X}_j\}\rangle /2
-\langle\hat{X}_i\rangle\langle\hat{X}_j\rangle$.
The CM $\sig$ of an arbitrary $N$-mode Gaussian state can be
written as follows in terms of $2 \times 2$ submatrices
\begin{equation}
\label{sigma}
 \sig =
\left(\begin{array}{ccc}
\sig_{1} & \cdots & \eps_{1N} \\
\vdots & \ddots & \vdots \\
\eps_{1N}^T & \cdots & \sig_{N} \\
\end{array}\right) \; .
\end{equation}
Symplectic operations ({\em i.e.}~belonging to the group
$Sp_{(2N,\R)}= \{S\in SL(2N,\R)\,:\,S^T\gr\Omega S=\gr\Omega\}$)
acting by congruence on CMs in phase space, amount to unitary
operations on density matrices in Hilbert space. Any $N$-mode
Gaussian state can be transformed by symplectic operations in its
Williamson diagonal form \cite{williamson36}
 $\gr\nu$, such that $\gr{\sigma}= S^T \gr{\nu} S$,
with $\gr{\nu}=\,{\rm diag}\,\{\nu_1,\nu_1,\ldots\nu_N,\nu_N\}$.
The quantities $\nu_i\ge 1$ are the symplectic eigenvalues of
$\gr{\sigma}$ \cite{extremal}.

We define the {\em symplectic rank} $\aleph$ of a CM as the number
of its symplectic eigenvalues different from $1$, corresponding to
the number of non-vacua normal modes. Any {\em pure} state has
$\aleph=0$. The CM $\sig^p$ of any $N$-mode pure Gaussian state
satisfies the matrix identity \cite{GEOF}  $- \gr\Omega\ \sig^p\
\gr\Omega\ \sig^p = \id$. As a consequence of the Schmidt
decomposition, applied at the CM level \cite{holewer} for the
bipartition $i|(1,\ldots,i-1,i+1,\ldots,N)$, any $(N-1)$-mode
reduced CM of the CM $\sig^p$ of a $N$-mode pure Gaussian state
has symplectic rank $\aleph = 1$.


\paragraph{Minimal number of parameters.\,{\em
---}} Adopting the above definition of generic entanglement, we prove now the
main
\\ \noindent{\em Proposition 1:} {A generic-entangled $N$-mode
pure Gaussian state is described, up to local symplectic (unitary)
operations, by $N(N-1)/2$ independent parameters.}


\noindent {\em Proof.} Let us start with a $N$-mode pure state,
described by a CM $\sig^p \equiv \sig$ with all single-mode blocks
in diagonal form: we can always achieve this by local single-mode
Williamson diagonalizations in each of the $N$ modes. Let
$\sig^{\backslash 1}\equiv \sig_{2,\cdots,N}$ be the reduced CM of
modes $(2,\dots,N)$. It can be diagonalized by means of a
symplectic $S_{2,\cdots,N}$, and brought thus to its Williamson
normal form, characterized by a symplectic spectrum
$\{a,1,\cdots,1\}$, where $a = \sqrt{\det{\sig_1}}$. Transforming
$\sig$ by $S=\id_1 \oplus S_{2,\cdots,N}$, brings the CM into its
Schmidt form, constituted by a two-mode squeezed state between
modes $1$ and $2$ (with squeezing $a$), plus $N-2$ vacua
\cite{holewer,botero}.

All $N$-mode pure Gaussian states are thus completely specified by
the symplectic $S_{2,\cdots,N}$, plus the parameter $a$.
Alternatively, the numbers of parameters of $\sig$ is also equal
to those characterizing an arbitrary mixed $N-1$ Gaussian CM, with
symplectic rank $\aleph=1$ ({\em i.e.}~with $N-2$ symplectic
eigenvalues equal to $1$). This means that, assigning the reduced
state $\sig_{2,\cdots,N}$, we have provided a complete description
of $\sig$. In fact,  the parameter $a$ is determined as the square
root of the determimant of the CM $\sig_{2,\cdots,N}$.

We are now left to compute the minimal set of parameters of an
arbitrary mixed state of $N-1$ modes, with symplectic rank
$\aleph=1$. While we know that for $N\ge4$ this number is equal to
$N(N-2)$ in general \cite{GEOF}, we want to prove that for
generic-entangled Gaussian resource states this number reduces to
\begin{equation}
\label{numero} \Xi_N={N(N-1)}/{2}\,.
\end{equation}
We prove it by induction. For a pure state of one mode only, there
are no reduced ``zero-mode'' states, so the number is zero. For a
pure state of two modes, an arbitrary one-mode mixed CM with
$\aleph=1$ is completely determined by its own determinant, so the
number is one. This shows that our law for $\Xi_N$ holds true for
$N=1$ and $N=2$.

Let us now suppose that it holds for a generic $N$, {\em i.e.}~we
have that a mixed $(N-1)$-mode CM with $\aleph=1$ can be put in a
standard form specified by $N(N-1)/2$ parameters. Now let us check
what happens for a $(N+1)$-mode pure state, {\em i.e.}~for the
reduced $N$-mode mixed state with symplectic rank equal to $1$. A
general way (up to LUs) of constructing a $N$-mode CM with
$\aleph=1$ yielding generic entanglement is the following: (a)
take a generic-entangled $(N-1)$-mode CM with $\aleph=1$ in
standard form; (b) append an ancillary mode ($\sig_N$) initially
in the vacuum state (the mode cannot be thermal as $\aleph$ must
be preserved); (c) squeeze mode $N$ with an arbitrary $s$ (one has
this freedom because it is a local symplectic operation); (d) let
mode $N$ interact couplewise with all the other modes, via a chain
of beam-splitters \cite{wolfito} with arbitrary transmittivities
$b_{i,N}$, with $i=1,\cdots,N-1$ \cite{footnotegen}; (e) if
desired, terminate with $N$ suitable single-mode squeezing
operations (but with all squeezings now {\em fixed} by the
respective reduced CM's elements) to symplectically diagonalize
each single-mode CM.

With these steps one is able to construct a mixed state of $N$
modes, with the desired rank, and with  generic (LU-invariant)
properties for each single-mode individual CM. We will show in the
following that in the considered states the pairwise quantum
correlations between any two modes are unconstrained. To conclude,
let us observe that the constructed generic-entangled state is
specified by a number of parameters equal to: $N(N-1)/2$ (the
parameters of the starting $(N-1)$-mode mixed state of the same
form) plus $1$ (the initial squeezing of mode $N$) plus $N-1$ (the
two-mode beamsplitter interactions between mode $N$ and each of the
others). Total: $(N+1)N/2 = \Xi_{N+1}$.  \hfill $\blacksquare$


\begin{figure}[t!]
\centering{\includegraphics[width=7cm]{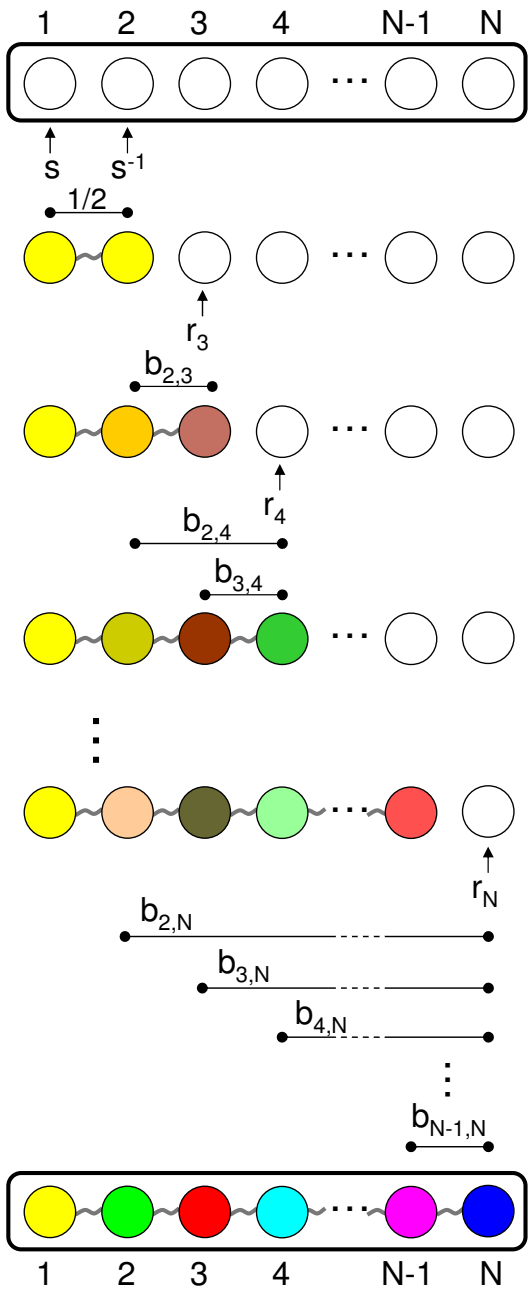}%
\caption{\label{figuro}(Color online) How to create a
generic-entangled  $N$-mode pure Gaussian state. White balls are
vacua, while each color depicts a different single-mode
determinant ({\em i.e.}~different degrees of local mixedness).
Vertical arrows denote single-mode squeezing operations, while
horizontal circle-ended lines denote beam-splitting operations
$b_{i,j}$ between modes $i$ and $j$. See text for details.}}
\end{figure}


\paragraph{Quantum state engineering.\,{\em ---}}
Following the ideas of the above proof, a physically insightful
scheme to produce generic-entangled  $N$-mode pure Gaussian states
can be readily presented (see Fig. \ref{figuro}). It consists of
basically two main steps: (1) creation of the state in the
$1|(N-1)$ Schmidt decomposition; (2) addition of modes and
entangling operations \cite{wolfito} between them. One starts with
a chain of $N$ vacua.

First of all (step 1), the recipe is to squeeze mode $1$ of an
amount $s$, and mode 2 of an amount $1/s$ ({\em i.e.}~one squeezes
the first mode in one quadrature and the second, of the same amount,
in the orthogonal quadrature); then one lets the two modes interfere
at a $50:50$ beam splitter. One has so created a two-mode squeezed
state between modes $1$ and $2$, which corresponds to the Schmidt
form of $\sig$ with respect to the $1|(N-1)$ bipartition. The second
step basically corresponds to create the most general mixed state
with $\aleph=1$, of modes $2,\cdots,N$, out of its Williamson
diagonal form. This task can be obtained, as already sketched in the
above proof, by letting each additional mode interact step-by-step
with all the previous ones. Starting with mode $3$ (which was in the
vacuum like all the subsequent ones), one thus squeezes it (of an
amount $r_3$) and combines it with mode $2$ via a beam-splitter
(characterized by a transmittivity $b_{2,3}$). Then one squeezes
mode $4$ by $r_4$ and lets it interfere sequentially both with mode
$2$ (with transmittivity $b_{2,4}$) and with mode $3$ (with
transmittivity $b_{3,4}$). This process can be iterated for each
other mode, as shown in Fig. \ref{figuro}, until the last mode $N$
is squeezed ($r_N$) and entangled with the previous ones via
beam-splitters with respective transmittivities $b_{i,N}$,
$i=2,\cdots,N-1$. Step 2 describes the distribution of the two-mode
entanglement created in step 1, among all modes.

The presented prescription enables to create a generic form (up to
LUs) of multipartite entanglement among $N$ modes in a pure
Gaussian state, by means of active (squeezers) and passive
(beam-splitters) linear optical elements. What is relevant for
practical applications, is that the state engineering is
implemented with minimal resources. Namely, the process is
characterized by  one squeezing degree (step 1), plus $N-2$
individual squeezings for step 2, together with $\sum_{i=1}^{N-2}
i = (N-1)(N-2)/2$ beam-splitter transmittivities, which amount to
a total of $N(N-1)/2 \equiv \Xi_N$ quantities. The optimally
produced Gaussian states can be readily implemented for $N$-party
CV communication networks \cite{network,review}.


\paragraph{Standard form.\,{\em ---}}
The special subset of pure $N$-mode Gaussian states emerging from
our constructive proof exhibits a distinct property: all
correlations between ``position'' $\hat q_i$ and ``momentum'' $\hat
p_j$ operators are vanishing. Looking at \eq{sigma}, this means that
such a generic-entangled pure Gaussian state can be put in a {\em
standard form} where all the $2\times2$ submatrices of its CM are
diagonal. The diagonal subblocks $\sig_i$ can be additionally made
proportional to the identity by local Williamson diagonalizations in
the individual modes. This standard form for generic-entangled
$N$-mode Gaussian states, as already mentioned, can be achieved by
{\em all} pure Gaussian states for $N=2$ \cite{2msform} and $N=3$
\cite{3modi}; for $N\ge4$, pure Gaussian states can exist whose
number of independent parameters scales as $N(N-2)$ \cite{GEOF} and
which cannot thus be brought in the $\hat q$-$\hat p$ block-diagonal
form. Interestingly, all pure Gaussian states in our considered
block-diagonal standard form are ground states of quadratic
Hamiltonians with spring-like interactions \cite{chain}.

Vanishing $\hat q$-$\hat p$ covariances imply that the CM can be
written as a direct sum $\sig^p = \gr V_Q \oplus \gr V_P$, when the
canonical operators are arranged as $\{\hat q_1,\ldots,\hat
q_N,\,\hat p_1,\ldots,\hat p_N\}$. Moreover, the global purity of
$\sig^p$ imposes $\gr V_P = \gr V_Q^{-1}$. Named
 $(\gr V_Q)_{ij}=v_{Q_{ij}}$ and
 $(\gr V_P)_{hk}=v_{P_{hk}}$,
this means that each $v_{P_{hk}}$ is a  function of the
$\{v_{Q_{ij}}\}$'s. The additional $N$ Williamson conditions
$v_{P_{ii}}=v_{Q_{ii}}$ fix the diagonal elements of $\gr V_Q$.
The standard form is thus completely specified by the off-diagonal
elements of the symmetric $N\times N$ matrix $\gr V_Q$, which are,
as expected, $N(N-1)/2 \equiv \Xi_N$. Proposition 1 acquires now a
remarkable physical insight: the structural properties of the
generic-entangled $N$-mode Gaussian states, and in particular
their bipartite and multipartite {\em entanglement}, are
completely specified (up to LUs) by the `two-point correlations'
$v_{Q_{ij}} = \langle \hat q_i \hat q_j \rangle$ between any pair
of modes. For instance, the entropy of entanglement between one
mode (say $i$) and the remaining $N-1$ modes, which is monotonic
in $\det \sig_i$ \cite{extremal}, is completely specified by
assigning all the pairwise correlations between mode $i$ and any
other mode $j \neq i$, as $\det \sig_i =1- \sum_{j \neq i} \det
\eps_{ij}$. The rationale is that entanglement in such states is
basically reducible to a mode-to-mode one. This statement,
strictly speaking true only for the pure Gaussian states for which
Proposition 1 holds, acquires a general validity in the context of
the modewise decomposition of arbitrary pure Gaussian states
\cite{holewer,botero}. This correlation picture breaks down for
mixed Gaussian states, where also classical, statistical-like
correlations arise.


\paragraph{Gaussian matrix product states.\,{\em
---}} As an application, let us consider Gaussian matrix product
states (GMPS),  defined as $N$-mode states obtained by taking a
fixed number, $M$, of infinitely entangled ancillary bonds
[Einstein-Podolski-Rosen (EPR) pairs] shared by adjacent sites, and
applying an arbitrary $2M \rightarrow 1$ Gaussian operation ${\cal
P}^{[i]}$ on each site $i=1,\ldots,N$. The projections ${\cal
P}^{[i]}$ can be described in terms of isomorphic $(2M+1)$-mode pure
Gaussian states with CM $\gr\gamma^{[i]}$, the {\em building blocks}
\cite{gvbs}.

It is conjectured that all pure $N$-mode Gaussian states can be
described as GMPS. Here we provide a lower bound on the number $M$
of ancillary bonds required to accomplish this task, as a function
of $N$. We restrict to ground states of harmonic chains with
spring-like interactions. With a simple counting argument, the total
number of parameters of the initial chain of building blocks should
be at least equal to that of the target state, {\em
i.e.}~$N(2M+1)(2M)/2 \ge N(N-1)/2$ which means $M \ge {\rm IntPart}
[(\sqrt{4N-3}-1)/4]$. This implies, for instance, that to describe
generic states with at least $N>7$ modes, a single EPR bond per site
is no more enough (even though the simplest case of $M=1$ yields
interesting families of $N$-mode GMPS for any $N$ \cite{gvbs}). The
minimum $M$ scales as $N^{1/2}$, diverging in the field limit $N
\rightarrow \infty$. As infinitely many bonds would be necessary
(and maybe not even sufficient) to describe generic infinite
harmonic chains, the matrix product formalism is probably not
helpful to prove or disprove area law statements for critical bosons
(complementing the known results for the non-critical case
\cite{area}), which in general do not fall in special subclasses of
finite-bonded GMPS.

The matrix product picture however effectively captures the
entanglement distribution in translationally invariant $N$-mode
harmonic rings \cite{gvbs}. In this case the GMPS building blocks
are equal at all sites, $\gr\gamma^{[i]} \equiv \gr\gamma \
\forall i$, while the number of parameters \eq{numero} of the
target state reduces to the number of independent pairwise
correlations (only functions of the distance between the two
sites), which by our counting argument is $\Theta_N \equiv (N -
N\!\mod 2)/2$. The corresponding threshold for a GMPS
representation becomes $M \ge {\rm IntPart}
[(\sqrt{8\Theta_N+1}-1)/4]$. As $\Theta_N$ is bigger for even $N$,
so it is the resulting threshold, which means that in general a
higher number of EPR bonds is needed, and so more entanglement is
inputed in the GMPS projectors and gets distributed in the target
$N$-mode Gaussian state, as opposed to the case of an odd $N$.
This clarifies why nearest-neighbour entanglement in ground states
of pure translationally invariant $N$-mode harmonic rings (which
belong to the class of states characterized by Proposition 1) is
frustrated for odd $N$ \cite{frusta}.


\paragraph{Conclusions.\,{\em ---}} In this Letter we studied pure $N$-mode
Gaussian states of CV systems, aimed to eradicate the minimal
number of degrees of freedom responsible for the generic nonlocal
features of the states. We showed that a crucial subclass of such
states, employed as typical resources in CV quantum information,
can be put in a standard form described (up to local unitaries) by
$N(N-1)/2$ parameters, corresponding to the ground states of a
quadratic Hamiltonian with spring-like couplings. This form
encompasses {\em all} pure Gaussian states for $N\le3$. We
operationally related these parameters with the active and passive
transformations needed to prepare the state. In general we
interpreted those degrees of freedom as the two-point correlations
between any pair of modes, which are thus responsible for the
structure of {\em generic entanglement} in Gaussian states.
It would be worth to
investigate the exotic properties of multipartite entanglement
arising in Gaussian states which {\em cannot} be brought in the
standard form described here for $N \ge 4$.
\\ \indent
I thank A.~Serafini, N.~Schuch, M.~M.~Wolf for crucial
clarifications, F. Illuminati for his wise advice, J.~Eisert,
M.~Ericsson, N.~Linden and J.~I.~Cirac for fruitful exchanges, and
M.~Plump for inspiring contacts.


\begin{thebibliography}{99}
\bibitem{footnotegen} Squeezings and beam-splitters are basic entangling tools in CV systems. For $N \ge 4$, steps (c) and (d)
should be generalized to arbitrary one- and two-mode symplectic
transformations to achieve all possible Gaussian states.
\end{thebibliography}
\end{document}